**Technical Conference Paper**

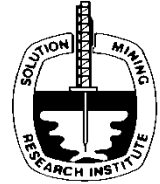

# Tensile Effective Stresses

# in Hydrocarbon Storage Caverns

**Hippolyte Djizanne and Pierre Bérest**
LMS, Ecole Polytechnique, Palaiseau, France

**Benoît Brouard**
Brouard Consulting, Paris, France

# TENSILE EFFECTIVE STRESSES IN HYDROCARBON STORAGE CAVERNS

Hippolyte Djizanne[1], Pierre Bérest[1], Benoît Brouard[2]

[1]LMS, Ecole Polytechnique, Palaiseau, France

[2]Brouard Consulting, Paris, France

**Abstract**

The "no-tensile effective stress" criterion is discussed. It is proven that effective tensile stresses can be generated at a cavern wall after a rapid increase or decrease in pressure. The Etzel K-102 test, performed in Germany more than 20 years ago, is revisited using the notion of effective tensile stresses.

**Key words:** Effective Stresses, Fracturing, Cavern Tightness

**Introduction**

Numerical computations allow for assessing the structural stability of salt caverns. Stresses, strains and volume changes can be computed and compared to safety criteria. The following criteria are commonly considered:

(1) no, or small, dilatant zone;

(2) limited volume loss and volume loss rate;

(3) limited subsidence;

(4) no, or small, tensile zone; and

(5) no, or limited, *effective* tensile zone.

The three first criteria have been discussed extensively in the literature. The two last criteria are discussed here.

**No Tensile Zone**

Tensile zones must be avoided, as the tensile strength of salt is low. Assuming zero tensile stress is on the safe side. Large tensile stresses lead to the creation of fractures. Deep fractures may be the origin of roof or wall spalling. In a salt cavern, stresses in the rock mass generally are compressive, and no tensile stresses appear at the cavern wall. Outstanding exceptions can be found, as noted and described below.

- When the cavern fluid pressure is low and the cavern profile exhibits non-convex portions — An example of this is described in Nieland and Ratigan, 2006.
- When the cavern pressure abruptly drops in a gas cavern — Rapid gas depressurization leads to gas cooling that is followed by slow gas warming when the cavern is kept idle. The decrease in gas temperature depends upon the relative withdrawal rate (in %/day), and upon cavern size and shape. Gas cooling results in the onset of additional tensile stresses at the cavern wall. If these additional tensile stresses are larger than the compressive stresses resulting from the difference between geostatic pressure and cavern gas pressure, the overall stresses at cavern wall are tensile, and may generate fractures or cracks[see, for example, Bauer and Sobolik (2009), Staudtmeister and Zapf (2010), Karimi-Jafari et al. (2011), Rokahr et al. (2011); Bérest et al. (2012), Leuger et al. (2012), and Lux and Dresen (2012).]

However, in most cases, the depth of penetration of these fractures is small, and they are perpendicular to the cavern wall. The distance between two parallel fractures becomes larger when fractures penetrate deeper into the rock mass, as some fractures stop growing. Salt slabs are created with boundaries formed by the opened fractures. As long as the depth of penetration of the fractures remains small, these slabs remain strongly bonded to the rock mass. It is believed that, in many cases, their weights are not large enough to allow them to break off at the cavern wall (Pellizzaro et al., 2011; Bérest et al., 2012).

- In ventilation mine shafts — "Thermal" fractures can be observed in unlined shafts (Wallner and Eickemeier, 2001; Zapf et al., 2012) as, in winter, shaft walls are cooled down over several months by ventilation air. Fractures are mainly horizontal, as vertical stresses at the cavern wall are less compressive than the natural geostatic stresses, due to the redistribution of viscoplastic stresses (see the Appendix).

**No Effective Tensile Zone**

**<u>Definition</u>**

The notion of "effective stress" applies to any porous rock, and it is commonly used in Reservoir Engineering. The effective stress equals the actual stress (Compressive stresses are negative.) plus the pressure of the fluid in the pores of the rock. Whether this notion applies to rock salt is a question still open to discussion, as salt permeability and porosity are exceedingly low. However, at a cavern wall, the effective stress simply is the actual stress plus the cavern pressure (Brouard et al., 2007). For simplicity, consider the case of an axi-symmetrical cavern. At cavern wall, three actual stresses must be considered: the normal stress, the tangential stress and the circumferential stress. In a perfectly cylindrical cavern, the tangential stress is the vertical stress. By definition, the effective normal stress is zero, as the actual normal stress is equal to minus the cavern pressure. The two other effective stresses may be positive (tensile) or negative (compressive).

It generally is accepted that when the effective stress is larger than a certain positive quantity, often called the rock tensile strength, or $T$, hydro-fracturing is possible. The related criterion can be written:

$$\sigma_{\min} + P < T \qquad (1)$$

where $\sigma_{\min}$ is the least compressive stress (Compressive stresses are negative.), and $P$ is the cavern pressure. When this criterion is not met, micro-fracturing occurs, permeability drastically increases, and salt softens. In some cases, discrete fractures appear (Bérest et al., 2001a,b; Rokahr et al., 2003 ; Düsterloh and Lux, 2012).

The tensile strength of salt, or $T$, is a couple of MPa or so. Selecting $T = 0$ (to be on the safe side), the criterion can be stated simply: The effective stress must be negative — i.e., "no tensile effective stress" must exist:

$$\sigma_{\min} + P < 0 \qquad (2)$$

In the following, several examples involving tensile effective stresses are discussed.

### Hydrofracturing

When performing a hydraulic fracturing test in a borehole, fluid is injected in the well and fluid pressure in the borehole is increased to a figure slightly higher than the geostatic pressure to create a fracture. At the beginning of the test, the injection rate is proportional to the rate of pressure increase; the ratio between these two quantities is the well compressibility. At some point, when the fluid pressure is high enough, compressibility increases drastically, a clear sign of the creation of a fracture or of a severe increase in permeability (Fokker, 1995; Bérest et al., 2001a,b; Rokahr et al., 2003; Lux et al., 2006). Finally, when the pressure reaches a maximum, a discrete fracture is created, and the pressure drops even when fluid is injected in the borehole. This maximum pressure is deemed to equal the least-compressive natural stress (plus rock strength). Such tests are performed routinely to assess in-situ stresses in salt formation [see Wawersick and Stone (1989), Schmidt (1993), Durup (1994), Rummel et al. (1996), Staudmeister and Schmidt (2000), and Doe and Osnes (2006)].

Instead of brine, gas sometimes is used during fracture (frac) tests to assess the tightness of the salt formation. This issue has been investigated in the special context of nuclear waste disposal: corrosion of waste packages generates gas; in a closed repository, gas pressure builds. To assess safety adequately, the risk of fracturing the salt formation must be discussed (see,for example, Popp et al., 2007).

The maximum pressure reached during a fractest provides an upper bound of the minimum geostatic stress. However, strictly speaking, such a simple interpretation is valid only when rock mechanical behavior is purely elastic. In the case of rock salt, interpretation may be more complicated, as the state of stress at the cavern wall is a function of the geostatic stress *and* of cavern pressure history.

When cavern pressure history is taken into account to compute stress distribution at a cavern wall, unexpected results can be found, as was mentioned first by Wawersik and Stone (1989). Tensile effective stresses may appear even when cavern pressure is quite low. Fracturing may appear when cavern pressure is smaller than geostatic stress. Examples are provided below.

### Tensile Effective Stress at a Gas-Cavern Wall (GasFrac)

It is commonly accepted that actual tangential stresses can be tensile at the wall of a gas cavern experiencing a severe pressure drop, because thermal tensile stresses are created. In such a case, it is obvious that effective tangential stresses are tensile, too. However, even when no tensile *actual stress* appears, tensile *effective stress* may be present. Examples can be found in Bauer and Sobolik (2009), Staudtmeister and Zapf (2010), Karimi-Jafari et al. (2011) and Rokahr et al. (2011). More recently, Lux and Dresen (2012) analyzed high-frequency, cycled storage gas and noted that **"**… a repeated situation can be observed where one of the tangential principal stresses at the cavern wall is smaller than the cavern pressure....This thermodynamic induced stress state demands special consideration**"** (p.378).

### Tensile Effective Stresses at a Brine-Filled Cavern Wall

It should be noted that such tensile effective stresses exist even in a cavern when no thermal effect is considered. Consider the case of a cavern kept idle at a relatively low pressure over a long period of time (say, several years). It can be proven that during such a period, due to cavern creep closure, deviatoric stress decreases. Because normal stress at a cavern wall remains constant, tangential stresses progressively become less compressive. When, at the end of such an "idle" period, gas or brine is injected rapidly into the cavern, fluid pressure increases, and *additional* tensile stresses are created. In some cases, these stresses are large enough to make effective stresses become tensile, and there is a risk of fracturing (Brouard et al., 2007), even when fluid pressure is (significantly) smaller than geostatic pressure. A possible example of this is the Etzel test.

**THE ETZEL K-102 TEST**

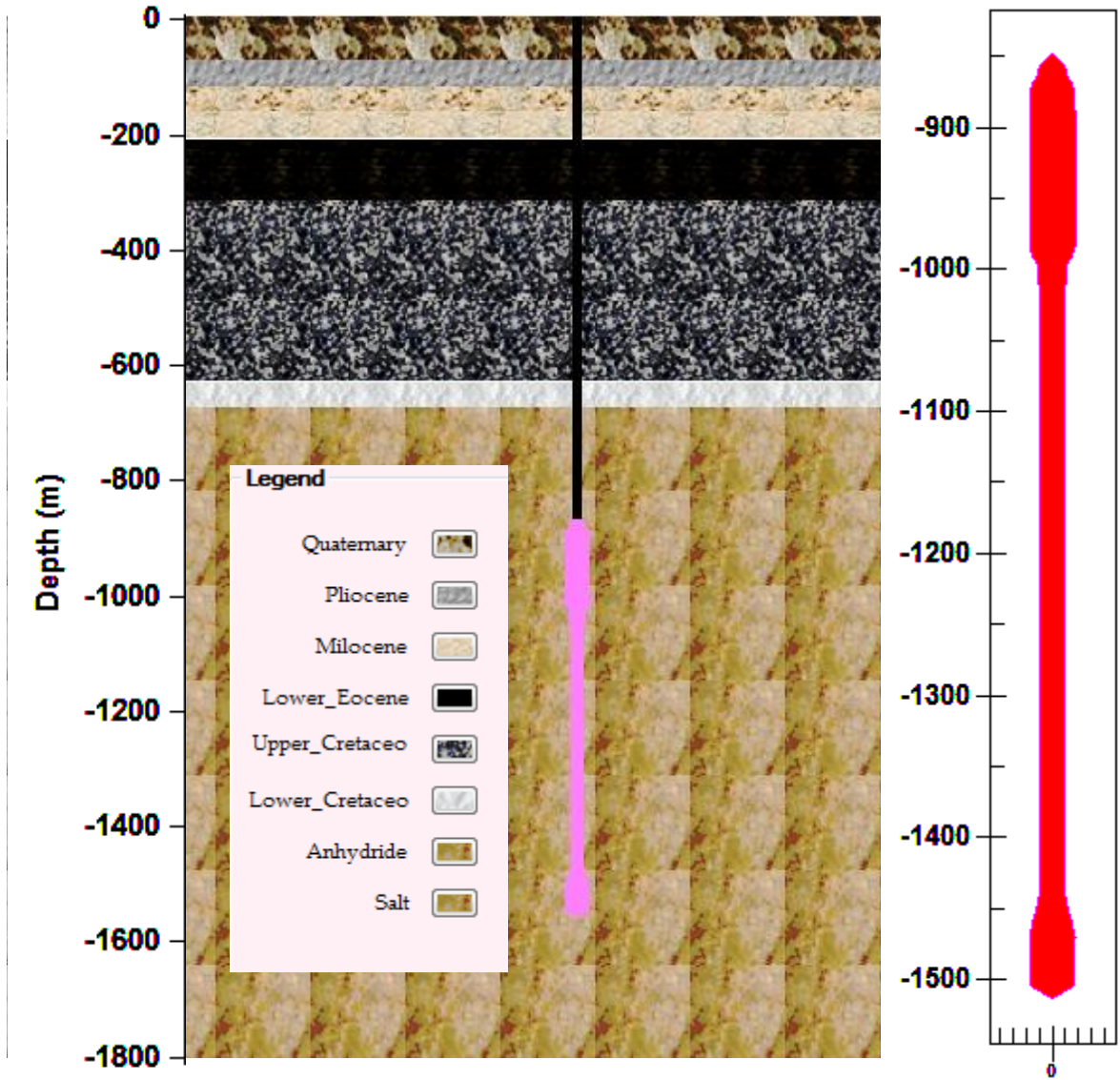

**Figure 1:** Simplified geometry and stratigraphy of the Etzel K-102 Cavern (After Rokahr et al., 2000).

Cavern K-102 was solution-mined mined between 1974 and 1978 (Figure 1). Oil was stored until April 1983. Oil then was removed and, later, the cavern was filled with brine — except from a short period in August 1987. Cavern pressure certainly was not constant during the 1978-1990 period (Pressure built up when the cavern was closed, due to liquid warming and cavern creep closure, and the cavern was vented periodically). A maximum brine-related head pressure of 6.95 MPa was reached at the end of February 1988; the associated gradient at casing-shoe depth was about $G$ = 0.0205 MPa/m.

In 1990, the Consortium Druckaufbautest decided to start a pressure build-up test on the Etzel K-102 cavern (Rokahr et al., 2000, from which the following description is drawn). In April 1989, the Etzel K-102 cavern volume was $V$ = 233,000 m$^3$. The casing-shoe depth was 827.7 m, the depth of the cavern roof was 850 m, and the cavern height was 622 m. The objective of the test was to obtain in-situ data as a basis for defining a cavern abandonment procedure. At that time, the state of the art was that cavern pressure could be increased to a high figure (a pressure gradient of $G$ = 0.032 MPa/m at casing-shoe depth) provided that pressure build-up rate was slow enough.

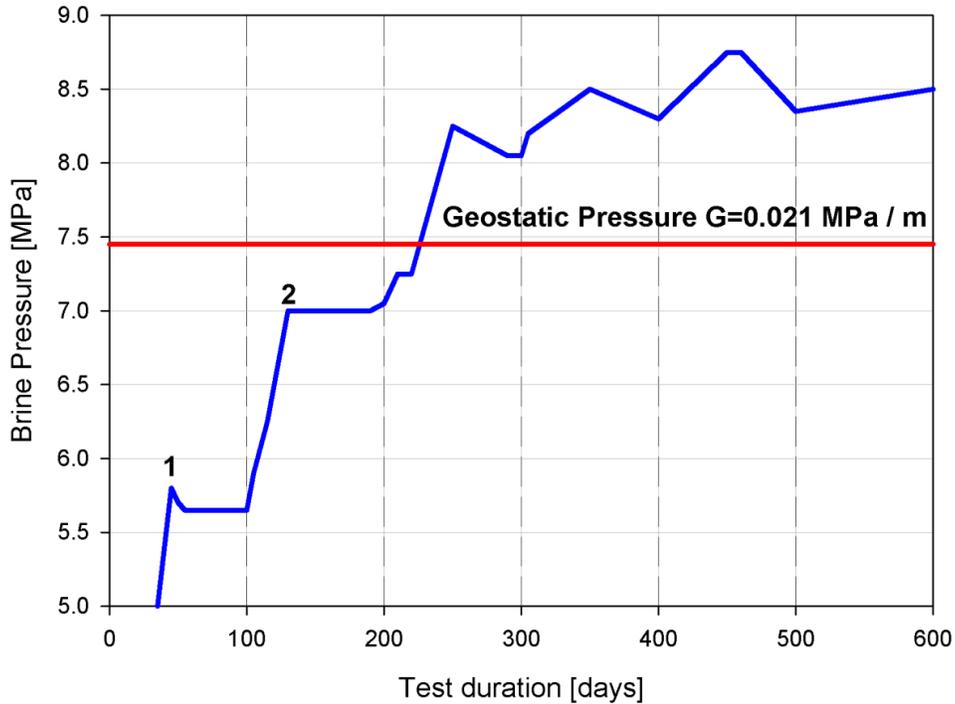

**Figure 2:** Wellhead Pressure Evolution During the Etzel K-102 Test (After Rokahr et al., 2000). Wellhead brine pressure is nil at the beginning of the test (day zero).

The pressure gradient at casing-shoe depth was increased incrementally. The first step, at the end of which $G$ = 0.19 bar/m (0.019 MPa/m at casing-shoe depth), was reached after injecting approximately $\Delta V = 500 \text{ m}^3$ in $\approx 45$ days. At casing-shoe depth, cavern pressure increased by $\Delta P = 827.7 \times (0.019 - 0.012) = 5.79 \text{ MPa}$ (1, Figure 2). The cavern-compressibility coefficient can be inferred to be $\beta = \Delta V / V \Delta P = 3.7 \times 10^4 / \text{MPa}$. (This is a relatively low figure). The second step, or $G$ = 0.0205 MPa/m, was reached after 134.4 m³ were injected from day 100 to day 130 (approximately), 2, Figure 2. The compressibility coefficient during this step was $\beta = 4.7 \times 10^4 / \text{MPa}$, a significant increase when compared to the first step. During the third step, compressibility consistently increased. After 7 weeks, 179.5 m³ had been injected, and a gradient of $G$ = 0.0219 MPa/m was reached. (The apparent compressibility coefficient was $\beta = 5.7 \times 10^4 / \text{MPa}$). After pumping stopped, cavern pressure consistently decreased. During the fourth step, a maximum gradient of $G$ = 0.233 bar/m was reached; however, "from this point on, the pressure in the cavern dropped gradually despite continued brine pumping" (Rokahr et al., 2000, p. 92.)

It was clear that, although cavern pressure was lower than geostatic pressure, some kind of micro-fracturing or permeability increase had taken place. Before the test, it was believed that the geostatic gradient was $G_{geo}$ = 0.0241 MPa/m, a high figure in the context of a salt formation. Well logs from existing and new boreholes were re-assessed, and suggested a revised figure of $G_{geo}$ = 0.0204 to 0.0211 MPa/m. However, it was observed that cavern compressibility significantly increased during the second pressure build-up, before a gradient at the casing-shoe depth of $G$ = 0.0205 MPa/m was reached, suggesting that fractures or "secondary" permeability might have been created long before geostatic gradient was reached.

It is suggested in this paper that the origin of the increase in permeability(or micro-fracturing) observed during the test may be the stress redistribution in the rock mass during the 1978-1900 period, when cavern pressure was relatively low.

**A TENTATIVE ANALYSIS OF EFFECTIVE STRESSES DURING THE ETZEL K-102 TEST**

**Cavern Creation and Test Duration**

It was noted that cavern pressure evolution during the period from 1983 to 2000 was not known perfectly. It was decided to assume that during this entire period, the cavern pressure was halmostatic — i.e., the gradient at the casing-shoe depth was $G = 0.012$ MPa/m. This assumption does not reflect fully the actual condition reigning before the test; it was selected because it is simple and makes the onset of tensile effective stresses easier. In fact, the following pressure history was selected (Figure 3).

- Cavern creation occurs over 300 days; the pressure gradient at the casing shoe decreases from $G = 0.021$ MPa/m (geostatic) to G = 0.012 MPa/m (halmostatic).

- The cavern is kept idle for the period $a = 2700$ days, during which the pressure gradient is halmostatic. Shorter ($a = 600$ days) and longer ($a = 6000$ days) durations also were considered to assess the influence of this parameter.

- The first step of the test is $b = 45$ day-longin the reference case (as it was during the actual test). The other steps are as described in Figure 2. Shorter ($b = 4$ days) and longer ($b = 1000$ days) durations were also considered.

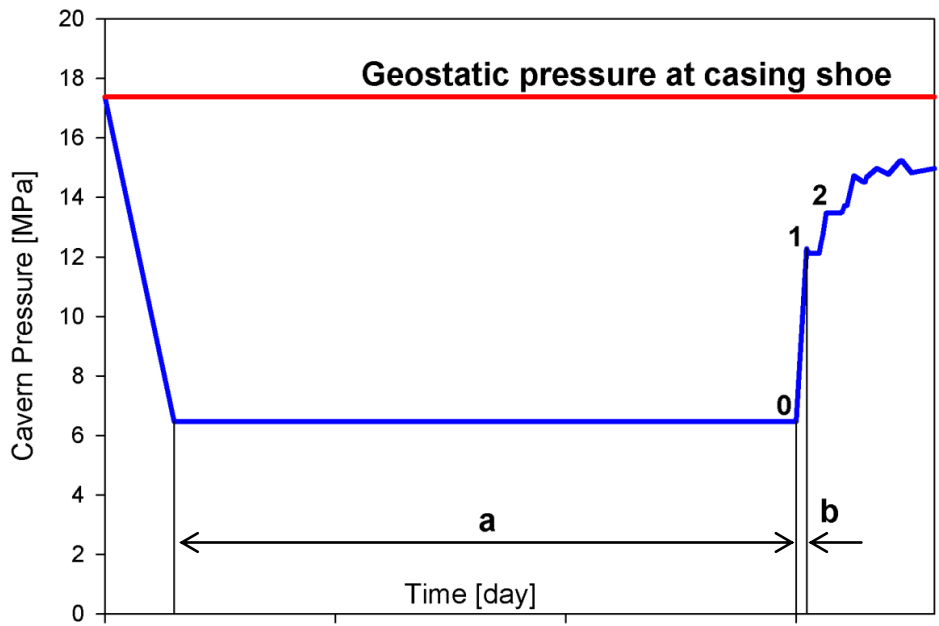

**Figure 3:** History of cavern pressure at casing-shoe depth (827.7 m). In the reference case, the cavern is leached out in 300 days; the waiting period is $a = 2700$-day long and the first pressure build-up is $b = 45$ day-long (duration of step 1).

## Constitutive Law

A Munson-Dawson constitutive law was selected. The values of the parameters of the Munson-Dawson law are providedin Table 1. In the reference case, the exponent of the power law is $n$ = 3.1, but higher values ($n$ = 4 and $n$ = 5) also were considered. Computations were performed using the Locas software. Computations were made usinga 43,000 elements mesh. A typical part of the mesh is shown on Figure 4.

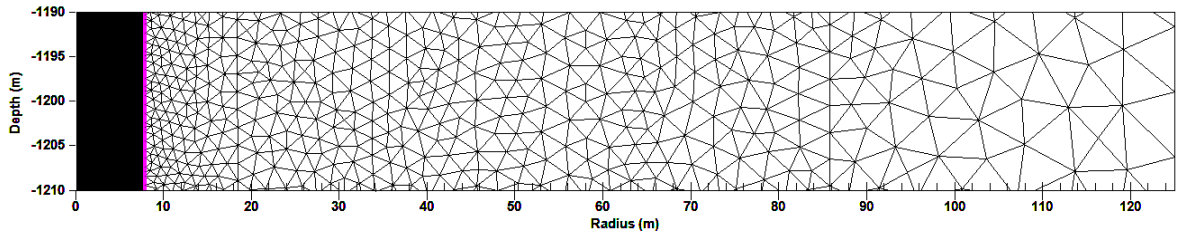

**Figure 4:** Computation mesh at a 1200-m depth (As a whole, 43 000 elements are used).

**Table 1.**     **Munson-Dawson Creep Law Parameters**

| Parameter | Unit | Value |
| --- | --- | --- |
| $A$ | /MPa^n-year | 0.64 |
| $n$ | - | 3.1 |
| $Q/R$ | K | 4100 |
| $m$ | - | 3 |
| $\alpha$ | - | -13.2 |
| $\beta$ | - | -7.738 |
| $K_0$ | /MPa^m | 7E-07 |
| $\delta$ | - | 0.58 |
| $c$ | /K | 0.00902 |

## Main Results, Reference Case

The results of the reference case are presented in Figures 5 and 6. Effective stresses at a point inside the rock mass are computed by adding to the actual stress the cavern pressure at the depth of the considered point in the rock mass. In other words, rock permeability is zero when the effective stress is compressive; it is infinite when the effective stress is tensile. *Such an assumption might be unrealistic.* Effective stresses and thicknesses of the tensile effective zones are represented at the end of steps 0, 1 and 2.The pressure gradients at casing-shoe depth are$G$ = 0.012 MPa/m at the end of step 0, $G$ = 0.019 MPa/m at the end of step 1 and $G$ = 0.0205 MPa/m at the end of step 2.  They are smaller than the geostatic gradient or $G_{geo}$ = 0.021 MPa/m.

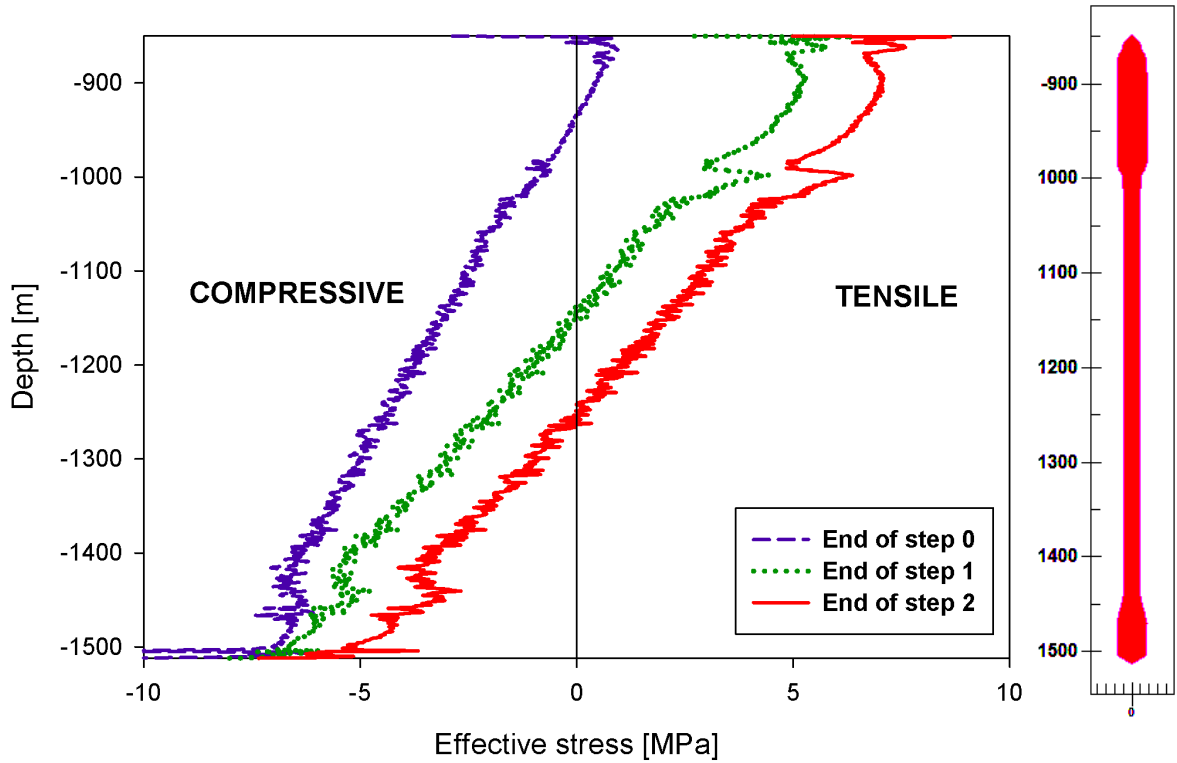

**Figure 5:** Distribution of the circumferential stress ($\sigma_{\theta\theta} + P$, in MPa) along the cavern wall at the end of steps 0, 1 and 2.

Figure 5 shows the Circumferential Effective Stress (CE*S*, or the effective stress in the direction perpendicular to the cross-sectional plane) at the cavern wall as a function of depth at the end of steps 0, 1 and 2. At the end of step 0 a small zone in which the CES is tensile can be observed in the upper part of the cavern. The tensile CES zone is much larger at the end of step 1, and larger still at the end of step 2.

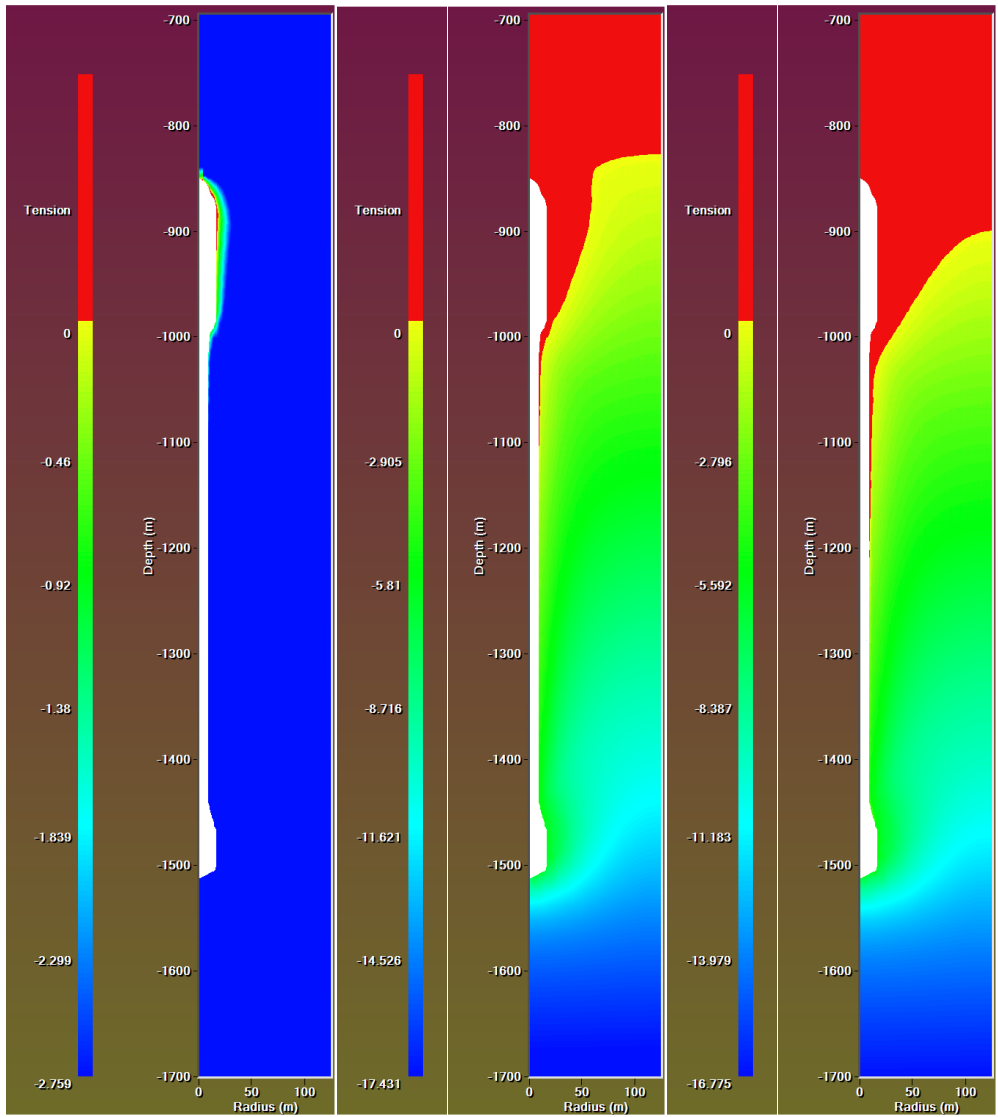

**Figure 6:** Contour plot of CES [in MPa] at the end of steps 0 (left), 1 (center) and 2 (right)

Figure 6 shows the thicknesses of the zonesin which the CESis tensile at the end of steps 0, 1 and 2. This zone is exceedingly small at the end of step 0; it originates in the non-convex shape of the cavern roof. At the end of steps 1 and 2, the tensile CES zone is much larger and reaches the 700-m deep salt roof.

**Radial distribution of the CES**

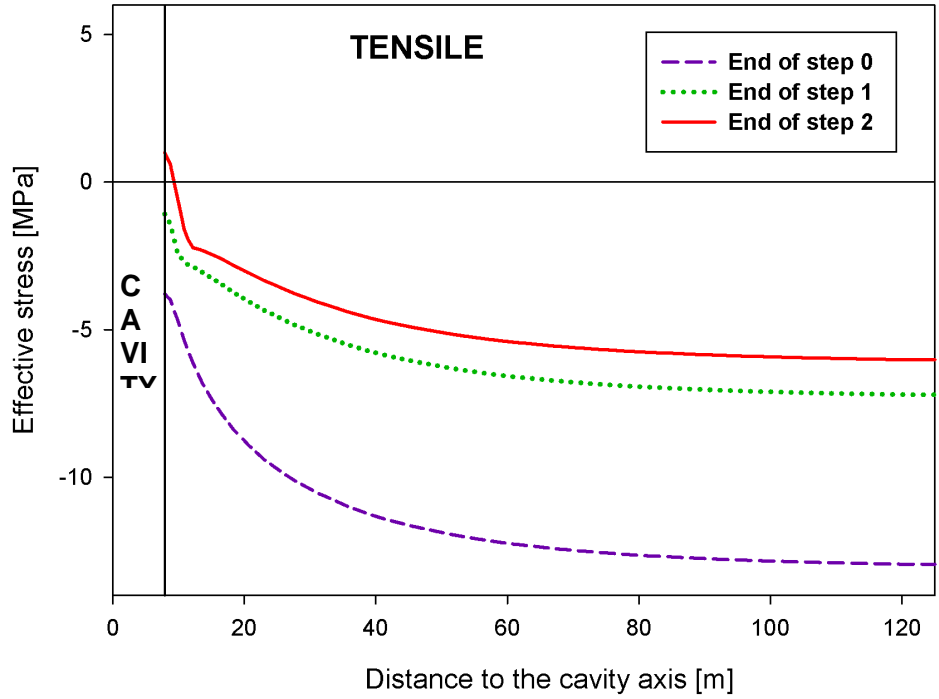

**Figure 7:** Radial distribution of effective stress at the end of steps 0,1 and 2 – As computed by LOCAS.

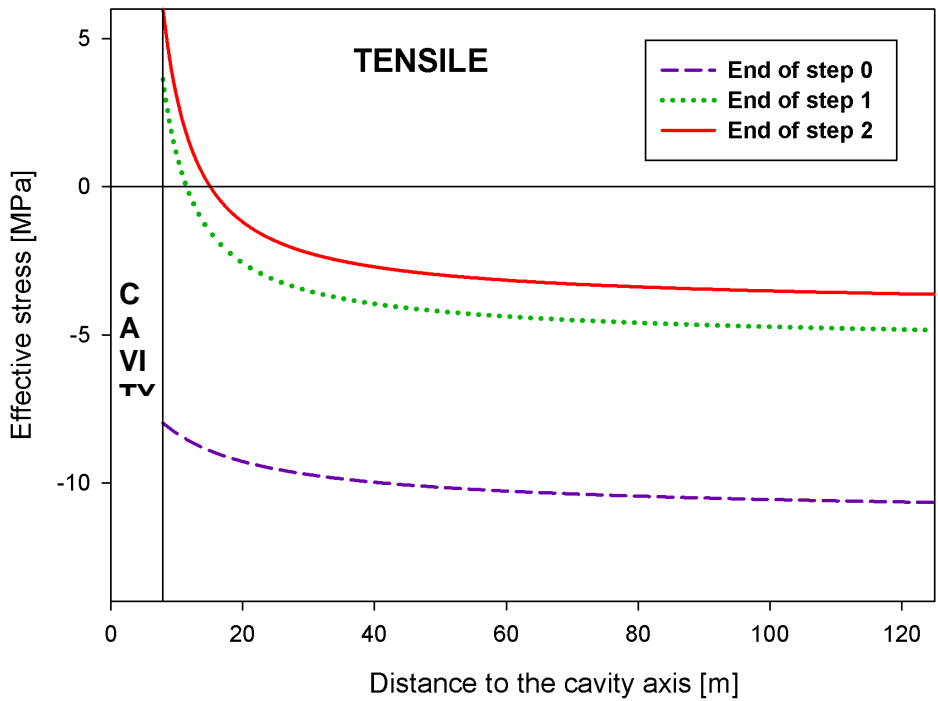

**Figure 8:** Radial distribution of effective stress at the end of step 0,1 and 2 – Closed-form solution; a simplified loading history is applied.

On Figure 7 is shown the distribution of the CES (or, $\sigma_{\theta\theta}+ P$) as a function of the distance from the cavern axis at the end of steps 0, 1 and 2. For comparison the same curves are given on Figure 8 when the closed-form solution (Appendix, Equation 4, no temperature is considered) is used. The closed-form solution assumes that steady-state was reached at the end of step 0, and that the response of the rock mass to the pressure build-up is elastic; differences between Figures 7 and 8 can be expected.

- Figure 7 proves that, at the end of step 0, steady-state has not been reached yet. Stress redistribution after an idle 2700-day long period is not completed, actual stress is less compressive than predicted by the steady-state solution, and the same can be said of the CES.

- Closed-form solution (Figure 8) assumes that, between step 0 and step 1, stress evolution is elastic: cavern pressure increases by 5.7 MPa and the CES at cavern wall increases by 2 x 5.7 MPa = 11.4 MPa. In the actual evolution (Figure 7), the pressure build-up phase is 45-day long, stresses are left enough time to redistribute, the deviatoric stress at cavern wall decreases, and the CES increases by 2.7 MPa (instead of 11.4 MPa in the closed-form solution)

- Closed-form solution predicts that, between step 1 and step 2, CES increases by 2 x 1.3 MPa = 2.6 MPa as cavern pressure increases by 1.3 MPa. In the actual evolution (Figure 7) the pressure build-up phase is 30-day long, during which viscoplastic stress redistribution takes place: CES increase is smaller. However, as cavern pressure increase is smaller from step 1 to step 2 than from step 0 to step 1, this redistribution is less effective.

**Variants**

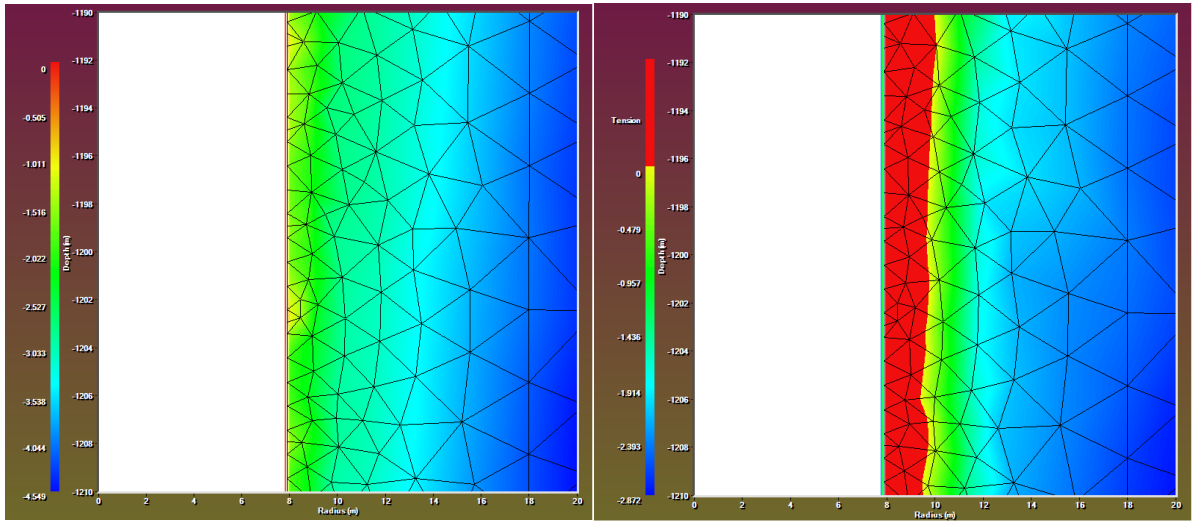

**Figure 9:**  Contour plot of CES ($\sigma_{\theta\theta}+ P$, in MPa) at the end of the step 2. Two pre-test waiting-period durations are considered: $a$ = 300 days (left) and $a$ = 5700 days (right). The reference duration is $a$ = 2700 days.

On Figure 9, two durations of the "waiting period" before the test, or $a$, are considered. The thickness of the zone in which CES are tensile at the end of step 2 is represented. This zone is thicker when the waiting period is longer. During the waiting period, at cavern wall, the normal stress is constant, and the deviatoric stress slowly decreases, making the circumferential stresses ($\sigma_{\theta\theta}$) less and less compressive. After 5700 days, tangential stresses are significantly less compressive than after 300 days. When pressure builds during the test, tensile tangential stresses are added; thus, the resulting effective stress is larger when the waiting period is longer.

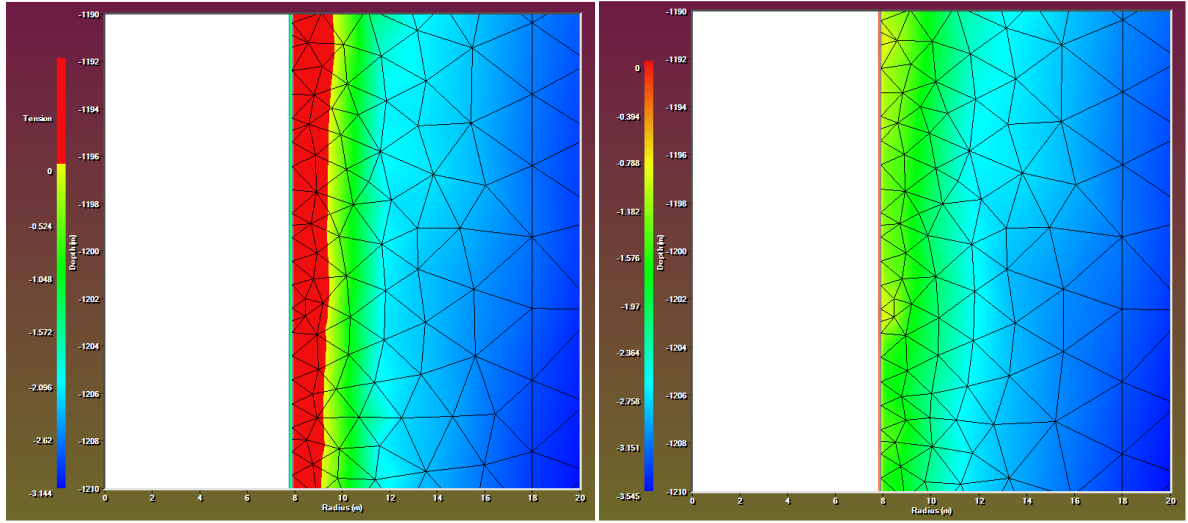

**Figure 10:** Contour plot of CES at the end of step 2. Two step-1 durations are considered: $b$ = 4 days (left) and $b$ = 1000 days (right). The reference duration is $b$ = 45 days.

On Figure 10, two durations of step-1 durations, or $b$, are considered. Step-2 duration remains unchanged. The thickness of the zone in which effective stresses are tensile at the end of step 2 is represented. This zone is thicker when the duration of step 1 is shorter. A slower pressure build-up rategives stresses more time to adjust.

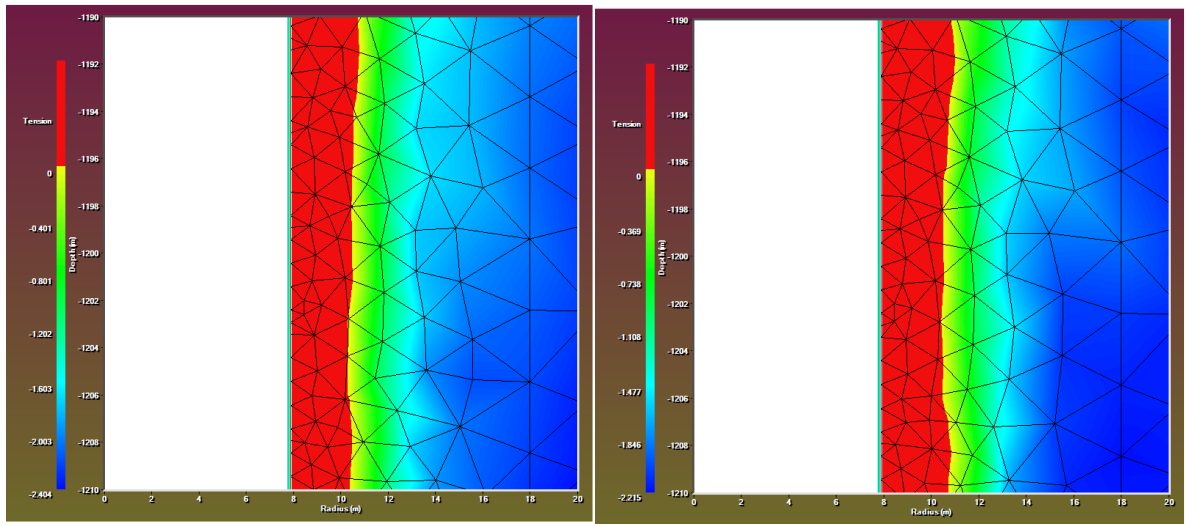

**Figure 10:** Contour plot of effective stress at the end of the step 2, illustrating the effect of the value of the stress exponent $n$. Two values are considered: $n$ = 4 and $n$ = 5. The reference case is $n$ = 3.1.

Figure 10 shows the effect of the value of stress exponent *n* on the thickness of the tensile CES zone. Two values are considered: *n* = 4, and *n* = 5. (The reference value is *n* = 3.1.) The zone in which effective stresses are tensile is thickest when *n* = 5. This result is more difficult to interpret than others. It is thought that two competing effects play a role. When *n* is larger, stress redistribution during the waiting period is slower (Brouard et al., 2007). Conversely, when *n* is larger, the value of the cavern pressure for which the tensile effective stress at cavern wall becomes tensile is lower (see Appendix, Equation 15)).

**Conclusions**

It was proven that, following a rapid pressure drop in a gas cavern or following a rapid pressure increase in a brine cavern, tensile effective stresses are generated at the cavern wall. These rapid changes may lead to micro-fracturing or to a permeability increase in a zone at the cavern wall. It is larger when the pressure change is faster and when the cavern has been kept idle for a long time before changing the pressure. In most cases, this zone is not very thick and does not seem to be a severe concern when cavern tightness is considered. However, it can be a concern when the distance between the top of the salt formation and the cavern roof is small. It must be noticed that the onset of tensile effective stresses must also be taken into account when interpreting frac tests.

**Acknowledgements**

This study was funded partially by the French *Agence Nationale de la Recherche* in the framework of the SACRE Project, which includes researchers from EDF, GEOSTOCK, PROMES (Perpignan), HEI (Lille) and Ecole Polytechnique (Palaiseau).

**APPENDIX**

Closed-form solutions often allow the main features of a complicated mechanical problem to be captured. Consider, for instance, the case of an idealized cylindrical salt cavern of radius *a*. The constitutive behavior of the salt mass is elastic-viscoplastic, and the viscoplastic part can be described by a Norton-Hoff or power law $\left(\dot{\varepsilon}^{vp} = A\sigma^n\right)$. Geostatic pressure at cavern depth is $P_\infty$. The cavern has been kept idle during a very long period of time, and steady-state creep closure has been reached. During this period, cavern pressure and temperature are constant and equal to $P_0$ and $T_0$, respectively. Then, cavern pressure is decreased rapidly, Figure A1 (left), or increased rapidly, Figure A1 (right), to $P_1$.

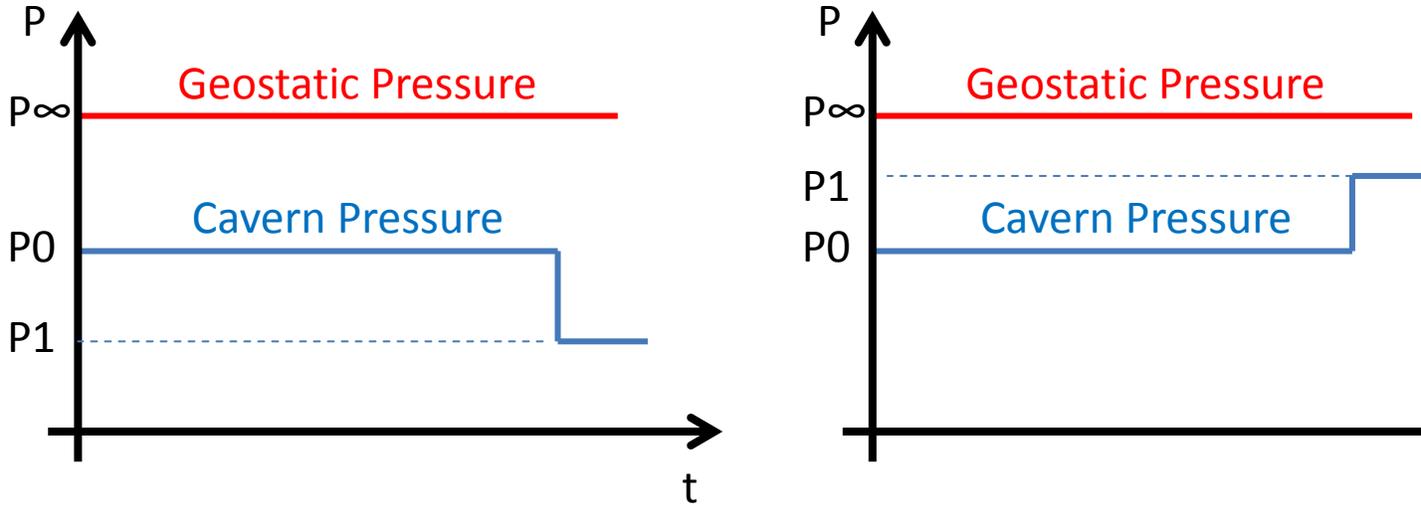

**Figure A1.** Sudden cavern-pressure decrease (left); Sudden cavern-pressure increase (right).

Pressure change is so fast that the response of the rock mass is thermo-elastic. In addition, the fluid temperature changes as a result of the change in fluid pressure. In a liquid storage cavern, this temperature change is exceedingly small. In a gas storage cavern, however, it cannot be neglected. The temperature change, or $\theta_g$, is largest when gas evolution is assumed to be adiabatic: $P_1/P_0 = \left(1 + \theta_g / T_0\right)^{1-1/\gamma}$, where $\gamma$ is the ratio between the constant-pressure and constant-volume heat capacities of the gas. The temperature change is positive when gas pressure increases, and is negative when gas pressure decreases. Additional stresses are created. The overall state of stresses can be written:

$$\sigma_{rr}(r) = -P_\infty + (P_\infty - P_0)\left(\frac{a}{r}\right)^{2/n} - \left(\frac{a}{r}\right)^2 \frac{E\alpha}{1-\nu}\int_a^r u\theta(u,t)\,du + (P_0 - P_1)\left(\frac{a}{r}\right)^2 \quad (3)$$

$$\sigma_{\theta\theta}(r) = -P_\infty + \left(1-\frac{2}{n}\right)(P_\infty - P_0)\left(\frac{a}{r}\right)^{2/n} + \left(\frac{a}{r}\right)^2 \frac{E\alpha}{1-\nu}\int_a^r u\theta(u,t)\,du - \frac{E\alpha\theta(r,t)}{1-\nu} - (P_0 - P_1)\left(\frac{a}{r}\right)^2 \quad (4)$$

$$\sigma_{zz}(r) = -P_\infty + \left(1-\frac{1}{n}\right)(P_\infty - P_0)\left(\frac{a}{r}\right)^{2/n} - \frac{E\alpha\theta(r,t)}{1-\nu} \quad (5)$$

where $r$ is the distance to the cavern axis, $n$ is the exponent of the power law, $E$ is the elastic modulus, $\nu$ is the Poisson's ratio, $\alpha$ is the coefficient of the rock-mass thermal expansion, and $\theta(r,t)$ is the temperature change in the rock mass generated by the initial fluid-temperature change by $\theta_g$.

At the cavern wall, or $r = a$, stresses can be written:

$$\sigma_{rr}(a) = -P_1 \tag{6}$$

$$\sigma_{\theta\theta}(a) = -P_\infty + \left(1 - \frac{2}{n}\right)(P_\infty - P_0) - \frac{E\alpha\theta_g}{1-\nu} - (P_0 - P_1) \tag{7}$$

$$\sigma_{zz}(a) = -P_\infty + \left(1 - \frac{1}{n}\right)(P_\infty - P_0) - \frac{E\alpha\theta_g}{1-\nu} \tag{8}$$

Two interesting cases can be considered.

1. Pressure drops in the cavern, $P_1 < P_0$, the cavern is filled with gas, the gas temperature drops by $\theta_g < 0$. — Note, first, that $\sigma_{zz} - \sigma_{\theta\theta} = (P_\infty - P_0)/n + (P_0 - P_1) > 0$: the vertical stress is less compressive than the circumferential stress ($\sigma_{\theta\theta}$). Furthermore, as $\theta_g < 0$, the vertical stress is tensile when the temperature drop is large enough, more precisely when

$$\left|\frac{E\alpha\theta_g}{1-\nu}\right| > \left(1 - \frac{1}{n}\right)P_0 + \frac{1}{n}P_\infty \tag{9}$$

Considera 800-m deep cavern, with $P_\infty = 17.6\,\text{MPa}$, $P_0 = 9.6\,\text{MPa}$, $E\alpha/(1-\nu) = 1\,\text{MPa/°C}$, and $n = 3$. Even a small temperature drop is able to generate tensile stresses. However, the onset of tensile stresses is more difficult in a deeper cavern, as both $P_0$ and $P_\infty$ are larger.

Obviously, effective stresses are more tensile than the actual stresses, as effective stresses are obtained by adding a positive quantity ($P_1$) to the actual stress. In fact, effective stresses can be tensile even when actual stresses are not. The condition for this is:

$$\left(1 - \frac{1}{n}\right)P_0 + \frac{1}{n}P_\infty - P_1 < \left|\frac{E\alpha\theta_g}{1-\nu}\right| < \left(1 - \frac{1}{n}\right)P_0 + \frac{1}{n}P_\infty \tag{10}$$

Note that, at the wall of a mine shaft, $P_0 = P_1 = 0$, and the condition (9) can be written:

$$\left|\frac{E\alpha\theta_g}{1-\nu}\right| > \frac{P_\infty}{n} \tag{11}$$

2. Pressure increases in the cavern, $P_0<P_1$—The cavern is filled with liquid (brine, for instance), and the liquid temperature increase is negligible. Now, the effective stresses at the cavern wall can be written:

$$\sigma_{rr}(a) + P_1 = 0 \tag{12}$$

$$\sigma_{\theta\theta}(a) + P_1 = -\frac{2}{n}P_\infty - 2\left(1-\frac{1}{n}\right)P_0 + 2P_1 \tag{13}$$

$$\sigma_{zz}(a) + P_1 = -\frac{1}{n}P_\infty - \left(1-\frac{1}{n}\right)P_0 + P_1 \tag{14}$$

Because $\sigma_{\theta\theta}(a) + P_1 = 2(\sigma_{zz}(a) + P_1)$, the circumferential and vertical effective stresses become tensile at the same time (when $P_1$ is increasing) —more precisely, when

$$P_1 - P_0 > \frac{1}{n}(P_\infty - P_0) \tag{15}$$

Consider, again, the 800-m deep cavern discussed above, with $P_\infty = 17.6 \, \text{MPa}$, $P_0 = 9.6 \, \text{MPa}$. Here, tensile effective stresses appear when the cavern pressure is larger than $P_1$ = 12.3 MPa, a relatively low figure.

Simple conclusions can be drawn from this closed-form solution, but a couple of remarks should be made first.

- It is assumed that steady-state viscoplastic behavior is reached before the cavern pressure is changed. Deviatoric stresses at cavern wall are quite large when the cavern is created; they slowly decrease when cavern pressure is kept constant to reach steady state. This process is quite slow (several years, or dozens of years). It is slower when cavern pressure is lower (i.e., when $P_\infty - P_0$ is larger) or when the exponent of the power law (*n*) is larger.

- It is assumed that, when cavern pressure changes, the response of the rock mass is elastic. This is true only when pressure change is extremely fast. In actuality, this response is not infinitely fast, and some viscoplastic evolution has time to take place.

For these reasons, numerical computations are needed to confirm the results of this simplified analysis.